\documentclass[fleqn,twoside]{article}
\usepackage{espcrc2,cite,graphicx,amsmath,amsfonts,amssymb,mathrsfs}
\usepackage[figuresright]{rotating}

\usepackage{amsmath}
\usepackage{amssymb}
\usepackage{epsfig}
\usepackage{graphicx}
\usepackage{cite}

\makeatletter
\renewcommand{\fnum@table}{\textbf{\tablename~\thetable}}
\renewcommand{\fnum@figure}{\textbf{\figurename~\thefigure}}
\makeatother

\newcounter{myenumi}

\renewcommand{\themyenumi}{\roman{myenumi}}
{\end{list}}

\newlength{\myem}
\newcounter{mysubequation}[equation]

\makeatletter
\renewcommand{\section}{\@startsection{section}{1}{0em}{-\baselineskip}%
{\baselineskip}{\normalfont\large\bfseries}}
\renewcommand{\subsection}%
{\@startsection{subsection}{2}{0em}{-0.7\baselineskip}%
{0.7\baselineskip}{\normalfont\bfseries}}
\makeatother

\newcommand{\bi}{\begin{itemize}}
\newcommand{\ei}{\end{itemize}}

\newcommand{\be}{\begin{equation}}
\newcommand{\ee}{\end{equation}}
\newcommand{\bea}{\begin{eqnarray}}
\newcommand{\eea}{\end{eqnarray}}

\newcommand{\ldm}{\Delta m_{31}^2}
\newcommand{\sdm}{\Delta m_{21}^2}
\newcommand{\deltacp}{\delta_{\mathrm{CP}}}
\newcommand{\stheta}{\sin^2(2 \theta_{13})}

\newcommand{\ie}{{\it i.e.}}

\newcommand{\eg}{{\it e.g.}}

\newcommand{\cf}{{\it cf.}}

\newcommand{\etc}{{\it etc.}}
\newcommand{\eq}{Eq.}

\newcommand{\fig}{Fig.}

\newcommand{\Ref}{Ref.}
\newcommand{\Refs}{Refs.}
\newcommand{\Sec}{Sec.}

\newcommand{\Tab}{Table}

\newcommand{\equ}[1]{\eq~(\ref{equ:#1})}
\newcommand{\figu}[1]{\fig~\ref{fig:#1}}

\newcommand{\NN}{\mathcal{N}}

\begin{document}
%%%%%%%%%%%%%%%%%%%%%%%%%%%%%%%%%%%%%%%%%%%%%%%%%%%%%%%%%%%%%%%%%%%%%
%%%%                     Title-page                              %%%%
%%%%%%%%%%%%%%%%%%%%%%%%%%%%%%%%%%%%%%%%%%%%%%%%%%%%%%%%%%%%%%%%%%%%%

\title{
\vspace*{-1cm}
\begin{flushright}
{\small SISSA 86/2005/EP}
\end{flushright}
\vspace*{0cm}
{\bf Testing mass-varying neutrinos with reactor experiments}}

\author{
{\large Thomas Schwetz}\address[SISSA]{{\it Scuola Internazionale Superiore di Studi Avanzati, 
       Via Beirut 2-4, 34014 Trieste, Italy\vspace*{-2mm}}}\thanks{E-mail:
{\tt schwetz@sissa.it}},
{\large Walter Winter}\address[IAS]{{\it
Institute for Advanced Study, School of Natural Sciences, Einstein Drive, Princeton, NJ 08540, USA}}\thanks{E-mail:
{\tt winter@ias.edu}}}

\begin{abstract}
\noindent {\bf Abstract}

\vspace*{0.2cm}
\noindent We propose that reactor experiments could be used to
constrain the environment dependence of neutrino mass and mixing
parameters, which could be induced due to an acceleron coupling to
matter fields.  There are several short-baseline reactor experiment
projects with different fractions of air and earth matter along the
neutrino path. Moreover, the short baselines, in principle, allow the
physical change of the material between source and detector.  Hence,
such experiments offer the possibility for a direct comparison of
oscillations in air and matter.
We demonstrate that for $\stheta \gtrsim 0.04$, two reactor
experiments (one air, one matter) with baselines of at least 1.5~km
can constrain any oscillation effect which is different in air and
matter at the level of a few per cent. Furthermore, we find that
using the same experiment while physically moving the material 
between source and detector improves systematics.

\vspace*{0.2cm}
\noindent {\it PACS:} 14.60.Pq \\
\noindent {\it Key words:} Neutrino oscillations, Matter effects, 
long-baseline experiments
\end{abstract}

\maketitle

\section{Introduction}

The concept of mass-varying neutrinos (MVNs) has been introduced by
imposing a relation between neutrinos and the dark energy of the
Universe~\cite{Hung:2000yg,Gu:2003er,Fardon:2003eh,Peccei:2004sz}
through a scalar field, the acceleron. Including the possibility of
acceleron couplings to matter fields implies that the neutrino
oscillation parameters in vacuum and a medium could be very
different~\cite{Kaplan:2004dq}, irrespective of the standard MSW
effect~\cite{Wolfenstein:1978ue,Mikheev:1985gs}. MVNs can have
substantial implications for neutrino phenomenology, for example in
the sun~\cite{Barger:2005mn,Cirelli:2005sg,Gonzalez-Garcia:2005xu} or
for cosmology and
astrophysics~\cite{Gu:2003er,Weiner:2005ac,Hung:2003jb}, and they have
been proposed as an explanation of the LSND
anomaly~\cite{Kaplan:2004dq,Zurek:2004vd,Barger:2005mh}.

Given existing data, precision tests of mass-varying effects are very
difficult, since the different experiments are conducted under
different conditions (different energies, baselines, matter
distributions, \etc). Practically, all evidence for neutrino
oscillations so far involves neutrino paths in (solar or
earth) matter, whereas no direct information is available on oscillation
parameters in vacuum or air. Moreover, the precise dependence of the
neutrino masses on the matter density through the acceleron coupling
is rather model dependent.
The only plausible laboratory for a direct comparison between neutrino
oscillations in different environments (\eg, matter and air) are
short-baseline experiments.
In this work, we consider the possibility to constrain MVNs with
reactor experiments, which are planned to measure the mixing angle
$\theta_{13}$~\cite{Minakata:2002jv,Huber:2003pm,Anderson:2004pk}.
Such experiments are particularly suitable for this purpose, because
one could think about building two very similar experiments with
different material along the baseline, or even about moving the
material along the baseline. In addition, reactor experiments suffer
very little from correlations among the oscillation parameters, and
new reactor experiments with identical near and far detectors will
have an excellent sensitivity to $\stheta$. Since the standard MSW
effect is irrelevant for such short baselines, any deviation of the
energy spectrum between matter and air can then be interpreted in
terms of MVNs.

%%%%%%%%%%%%%%%%%%%%%%%%%%%%%%%%%%%%%%%%%%%%%%%%%%%%%%%%%%%%%%
\section{General formalism}
%%%%%%%%%%%%%%%%%%%%%%%%%%%%%%%%%%%%%%%%%%%%%%%%%%%%%%%%%%%%%%

For the reactor experiments under consideration, the ordinary matter
potential is, to a first approximation, negligible because of the
short baseline. The oscillation probability reads in the expansion up
to second order in $\sin (2 \theta_{13})$ and $\alpha \equiv \sdm/\ldm$
(\cf, \eg, \Ref~\cite{Akhmedov:2004ny})
\begin{eqnarray}
1 - P_{\bar{e} \bar{e}}  & = &
\sin^2 (2 \theta_{13}) \, \sin^2 \Delta_{31} + \nonumber \\
& & + \, \alpha^2 \, \Delta_{31}^2 \, \sin^2 ( 2 \theta_{12}) \,,\label{equ:osc}
\end{eqnarray}
where $\Delta_{31} \equiv \ldm L/(4E)$. Thus, one can easily see that
short-baseline reactor experiments have an excellent sensitivity to
$\stheta$. In particular, since $\alpha^2 \simeq 10^{-3}$, one can
read off from this equation that for large $\stheta \gg 10^{-3}$ the
second term is practically negligible, and we can use this formula in
the two-flavor limit for analytical purposes.

Suppose that we have {\em any} non-standard effect on neutrino
oscillations on the Hamiltonian level which is different in air and
matter. In this case, we have the usual vacuum Hamiltonian
$\mathcal{H}_{\mathrm{vac}}$ in air, and some non-standard Hamiltonian
$\mathcal{H}_{\mathrm{mat}} = \mathcal{H}_{\mathrm{vac}} +
\mathcal{H}_{\mathrm{ns}}$ in matter. In general,
$\mathcal{H}_{\mathrm{ns}}$ is a Hermitian traceless\footnote{If it is
not traceless, it can be made traceless by the subtraction of a global
phase.} $n \times n$ matrix (for details, see \eg\
\Ref~\cite{Blennow:2005qj}). The Hamiltonian
$\mathcal{H}_{\mathrm{mat}}$ can then be re-diagonalized in order to
obtain the effective mixing angles $\tilde{\theta}_{13}$ and $\Delta
\tilde{m}^2_{31}$ in matter. In order to parameterize deviations
between matter (tilde) and air (no tilde) parameters, we define
\begin{eqnarray}
\delta_\Delta & = & \ldm - \Delta \tilde{m}^2_{31} \, , \nonumber \\
\delta_\theta & = & \theta_{13} - \tilde{\theta}_{13} \, .\label{equ:uparam}
\end{eqnarray}
For MVNs one has~\cite{Barger:2005mn,Gonzalez-Garcia:2005xu} 
\begin{equation}\label{equ:MVN}
\mathcal{H}_{\mathrm{mat}} = \frac{1}{2E} \, U 
\left[\hat m - M(\rho) \right]^\dagger 
\left[\hat m - M(\rho) \right]
U^\dagger \,,
\end{equation}
where $\hat m = \mathrm{diag}(m_i)$ is the diagonal matrix of the
neutrino masses and $U$ is the mixing matrix, both in the background
dominated environment (\eg, air), and $M(\rho)$ is the mass matrix
depending on the matter density $\rho$, which in general is
non-diagonal in the basis used in \equ{MVN}. Hence, one finds that in
the case of MVNs $\mathcal{H}_{\mathrm{vac}}$ and
$\mathcal{H}_{\mathrm{ns}}$ have the {\em same} energy dependence,
\ie, it is indeed the mass matrix which is modified. Note that
according to \equ{MVN}, $\mathcal{H}_{\mathrm{ns}}$ also depends on
the $m_i$.

In the following we will assume that the neutrinos propagate either
through air or through matter with constant density. This allows a
model-independent test of MVNs, since only the density difference
enters, and we need not to specify the detailed functional dependence
of $M_{ij}(\rho)$.
Furthermore, we assume that environment-dependent variations of
$\Delta m^2_{21}$ and $\theta_{12}$ can be neglected, and the two
parameters $\delta_\Delta$ and $\delta_\theta$ suffice to describe MVN
effects for reactor experiments under consideration. This assumption
can be justified by requiring the consistency of solar neutrino and
KamLAND data, due to the much higher matter densities in the sun. The
typical size of MVN parameters relevant for solar neutrinos
\cite{Barger:2005mn,Gonzalez-Garcia:2005xu} is about one order of
magnitude smaller than the sensitivity of short-baseline reactor experiments,
and in particular, if the bounds derived in
\Ref~\cite{Gonzalez-Garcia:2005xu} are applied, MVN effects on $\Delta
m^2_{21}$ and $\theta_{12}$ can be neglected (see also
\Sec~\ref{sec:models}). We stress that reactor experiments test {\em
different} MVN parameters than solar neutrino experiments.

Let us add, that for non-standard matter effects apart from MVNs (see
\Ref~\cite{Valle:1990pk} for a review) the energy dependence of
$\mathcal{H}_{\mathrm{ns}}$ may be different than the one of
$\mathcal{H}_{\mathrm{vac}}$. For instance, there may be a relative
factor $1/(2E)$ between $\mathcal{H}_{\mathrm{vac}}$ and
$\mathcal{H}_{\mathrm{ns}}$ leading to energy-dependent matter
parameters similar to the standard MSW matter effect (\cf,
\Ref~\cite{Blennow:2005qj}). In this case, the assumption of an
energy-independent effect can still be used as an approximation for
short-baseline reactor experiments (by using the peak energy 
of the spectrum) as long as the neutrino energy is
far away from the MSW resonance energy. This requirement is only violated
if the non-standard effects are more than two orders of magnitude
larger than the standard MSW matter effect.

%%%%%%%%%%%%%%%%%%%%%%%%%%%%%%%%%%%%%%%%%%%%%%%%%%%%%%%%%%%%%%%%%%
\section{Reactor experiment geometries}
%%%%%%%%%%%%%%%%%%%%%%%%%%%%%%%%%%%%%%%%%%%%%%%%%%%%%%%%%%%%%%%%%%
\label{sec:setups}

\begin{figure}[t]
\begin{center}
\begin{tabular}{c}
\includegraphics[width=0.9\columnwidth]{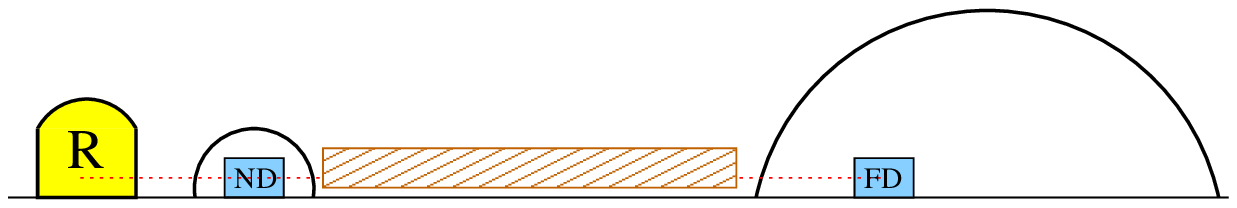} \\
(a) Mostly air \\[0.5cm]
\includegraphics[width=0.9\columnwidth]{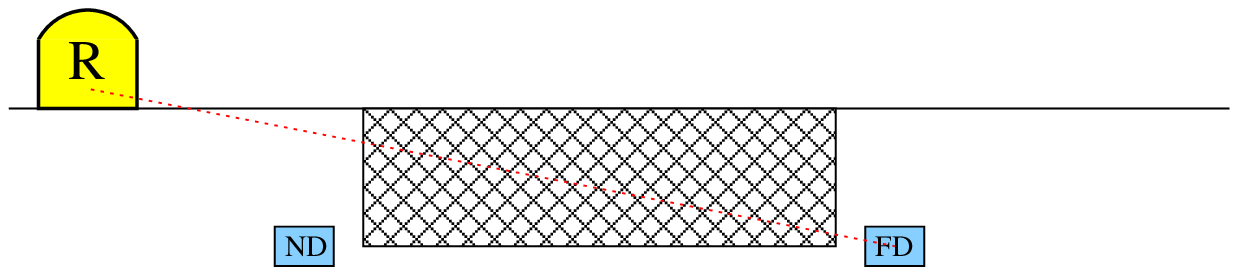} \\
(b) Flat terrain, underground \\[0.5cm]
 \includegraphics[width=0.9\columnwidth]{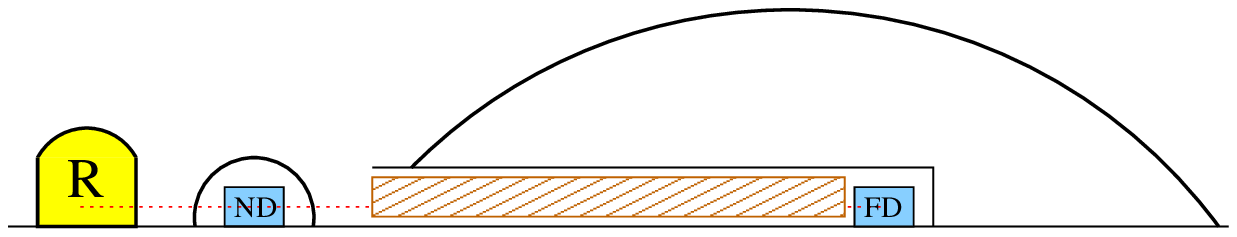} \\
(c) Hills, mostly underground
\end{tabular}
\caption{\label{fig:types} Schematic illustration of different types
  of reactor experiment terrain layouts and potential modifications
  for the test of non-standard matter effects (not to scale).  The
  filled boxes in (a) and (c) refer to additions of material in the
  far detector's line of sight, and the box in (b) refers to removing
  material in the far detector's line of sight after the primary
  operation period.}
\end{center}
\end{figure}

In principle, we expect to obtain the best MVN sensitivities by the
comparison of a matter to an air experiment. As we will see, using
reactor experiments for the test of MVNs is mainly a matter of
geometry. Some experiments have their baseline in air (such as
Double Chooz; see, \eg, \Refs~\cite{Zurek:2004vd,Barger:2005mh}), 
and others have it mostly in matter or mixed between
air and matter sections. In addition, moving detectors have been
discussed.  Currently, three major principle schemes of reactor
experiment layouts, which can be found in \figu{types}, are:
\begin{description}
\item[Type (a)] Baseline mostly in air. 
Example: Double Chooz~\cite{Ardellier:2004ui}.
\item[Type (b)] Baseline entirely in matter because of flat terrain. 
Example: Braidwood~\cite{Bolton:2005yd}, KASKA~\cite{Kuze:2005kz}.
\item[Type (c)] Baseline mostly in matter because of hills. 
Examples: Daya Bay~\cite{Cao:2005mh},  Angra~\cite{angra}.
\end{description}
One could now think about three different conceptual cases to test MVN
effects.  First, one could use existing (future) experiments with
baselines in air and matter and combine them, such as Double Chooz
or a larger reactor experiment in air combined with another reactor 
experiment in matter, or the
long-baseline beam experiments T2K or NO$\nu$A.  Second, one could
modify the existing experiments. In phase~I, the experiment runs
unmodified, and in a subsequent phase~II the material in the line of
sight of the far detector is changed. Examples of these modifications
can be found in \figu{types}. For type (a), the far detector's line of
sight could be covered with Earth or a rock wall. For type (b), a hole
in the ground would lead to air propagation in the far detector's line
of sight. And for type (c), the access tunnel to the far detector
could a priori be built in the line of sight of far detector which is
then covered with Earth or rock wall in phase~II. Alternatively, one
could drill an additional tunnel. As the last conceptual case, one
could use a vertically or horizontally moving detector and the
geometry of the site to change the material between source and
detector.

%%%%%%%%%%%%%%%%%%%%%%%%%%%%%%%%%%%%%%%%%%%%%%%%%%%%%%%%%%%%%%%%%%%%%%%%
\section{Analysis methods}
%%%%%%%%%%%%%%%%%%%%%%%%%%%%%%%%%%%%%%%%%%%%%%%%%%%%%%%%%%%%%%%%%%%%%%%%

For the experiment simulation, we use a version of the GLoBES
software~\cite{Huber:2004ka} with some modifications to simulate
non-standard physics~\cite{Blennow:2005yk,GLOBESnew}.  For the
experiments, we mainly use the simulations of the Reactor-II and
Double Chooz~\cite{Ardellier:2004ui} setups from
\Refs~\cite{Huber:2003pm,Huber:2004ug}.  Reactor-II is an abstract
background-free reactor experiment with identical near and far
detectors, a baseline $L_{\mathrm{far}}=1.7 \, \mathrm{km}$, and an
integrated luminosity of $8000 \, \mathrm{t \, GW \, yr}$. There are
now many proposals for an experiment similar to this setup (\cf,
\Ref~\cite{Anderson:2004pk}, and references therein).  For Double
Chooz, we use a background-free simulation with an integrated $60 \,
000$ unoscillated events, and a baseline $L_{\mathrm{far}}=1.05 \,
\mathrm{km}$, which should be a very good approximation for the actual
experiment. In addition, we will show some results in combination with
NO$\nu$A instead of a reactor experiment in matter, where we use the
simulation based upon \Refs~\cite{Huber:2002mx,Huber:2002rs} adjusted
to the proposal parameters~\cite{Ayres:2004js} ($L=810 \,
\mathrm{km}$, $12 \, \mathrm{km}$ off-axis, 30~kt TASD) with a running
time of five years in the neutrino mode.

We simulate the data for a given set of ``true'' oscillation
parameters~\cite{Fogli:2005cq,Bahcall:2004ut,Bandyopadhyay:2004da,Maltoni:2004ei}
$\ldm = 2.2 \, \cdot 10^{-3} \, \mathrm{eV}^2$, $\sin^2 2
\theta_{23}=1$, $\sdm = 8.2 \cdot 10^{-5} \, \mathrm{eV}^2$, $\sin^2 2
\theta_{12}=0.83$, and $\stheta=0.1$ somewhat below the CHOOZ
bound~\cite{Apollonio:2002gd}, $\deltacp=0$ (relevant for beams only),
and a normal hierarchy, and we assume no MVN effect in the ``data'',
\ie, true $\delta_\Delta=\delta_\theta=0$.  Then we perform a fit to
these simulated data, and we project the fit manifold onto the
$\delta_\Delta$-$\delta_\theta$ plane by marginalization of the
standard oscillation parameters. For this process, we impose external
precisions of 5\% for each $\sdm$ and
$\theta_{12}$~\cite{Bahcall:2004ut}, which should be realistic
estimates for the analysis time.  We do not include the wrong
hierarchy solution, because there is a good chance that the mass
hierarchy will be measured on the timescale we are discussing (see,
\eg, \Refs~\cite{Ayres:2004js,Huber:2004ug}).\footnote{Because reactor
experiments are insensitive to the mass hierarchy, the results do, in
principle, not depend on the mass hierarchy. However, the sign of the
$\delta_\Delta$-effects could change for a different mass hierarchy,
because $\delta_\Delta$ may come from a specific shift of one of the
mass eigenstates.}

%%%%%%%%%%%%%%%%%%%%%%%%%%%%%%%%%%%%%%%%%%%%%%%%%%%%%%%%%%%%%%%%%%%%%%%%
\section{Systematics treatment}
%%%%%%%%%%%%%%%%%%%%%%%%%%%%%%%%%%%%%%%%%%%%%%%%%%%%%%%%%%%%%%%%%%%%%%%%

Before we present our results, we need to discuss the reactor
experiment systematics in greater detail. We use a treatment similar
to \Ref~\cite{Huber:2003pm} for a reactor experiment with two
identical detectors, where the main systematics impact was identified
as an effective normalization error of about $0.8\%$. Here we consider
two different cases:
\begin{enumerate}
\item
One experiment with two phases: Phase~I mainly in matter, phase~II
mainly in air. The detectors are not only identical, but also
physically the same.
\item
Two different experiments, one mainly in air, the other mainly in
matter. The detectors are identical, but not physically the same.
\end{enumerate}
We write the $\chi^2$ for both cases as
\begin{eqnarray} 
\chi^2 &=& \sum_{x=m,a} \left[ \sum_i  
\frac{ [(1+b+b_x) N_i^x - O_i ]^2 }{O_i} \right. \nonumber \\
&& \qquad \left. + \left(\frac{b_x}{\sigma_{uc}}\right)^2 \right] + 
\left(\frac{b}{\sigma_c}\right)^2\, ,
\label{equ:chi2}
\end{eqnarray}
where $N_i$ ($O_i$) are the oscillated (unoscillated) event numbers in
energy bin $i$, the index $x$ labels the data taken in air ($a$)
and in matter ($m$), and one has to minimize \equ{chi2} with respect
to $b$, $b_a$, and $b_m$. The error $\sigma_{uc}$ corresponds to an 
uncorrelated normalization error between the air and matter data, whereas
$\sigma_c$ is fully correlated. The interpretation of $\sigma_{uc}$ and
$\sigma_c$ depends on the setup and will be discussed below.

Let us do some analytical estimations. First, we will see that the
systematics mainly affects the $\stheta$ measurement, which means that
we will be only interested in the $\stheta$ measurement in this
section. Second, for simplicity, we consider only a rate
measurement. Then the oscillated event rates can be written as $N^x =
(1 - S_x f_x) \NN$, where $S_x \equiv \sin^2 (2\theta_{13}^x)$,
\begin{equation}\label{equ:def}
f_x \equiv \left< \sin^2\frac{\Delta m^2_x L}{4E}\right> = \mathcal{O}(1) \, ,
\end{equation}
and $\NN \equiv O$ is the total number of events in one experiment.
Using $S_x, b, b_x \ll 1$, the $\chi^2$ can be linearized. Since we
are interested in the absolute error $\sigma_S$ on $S_a - S_m$, we
then minimize the linearized \equ{chi2} with respect to $b$, $b_a$,
and $b_m$, and compute $\sigma_S$ taking into account the
correlation matrix between $S_a$ and $S_m$ in order to obtain
\begin{equation}\label{equ:sigmau}
\sigma_S^2 \simeq
\left(\sigma_{uc}^2 + \frac{1}{\NN}\right)
\left(\frac{1}{f_a^2} + \frac{1}{f_m^2}\right) 
+
\sigma_c^2
\left(\frac{1}{f_a} - \frac{1}{f_m}\right)^2 \,.
\end{equation}
For case~2 (two separate experiments), $\sigma_{uc}$ corresponds to
the effective normalization error of each individual experiment, and
the second term of \equ{sigmau} can be neglected because it is usually
small compared to the first one. As one may expect, in the (poor)
statistics dominated regime $1/\sqrt{\NN} \gg \sigma_{uc}$, the error
scales as $1/\sqrt{\NN}$, whereas for large event numbers
$1/\sqrt{\NN} \ll \sigma_{uc}$ the error is limited by the
systematical uncertainty $\sigma_{uc}$ (\cf, also
\Refs~\cite{Sugiyama:2004bv,Sugiyama:2005ir,Huber:2004bh} for similar
considerations). For Reactor-II, we have $\NN \sim 630 \, 000$
unoscillated events and $\sigma_{uc} \simeq 0.8\%$, which means that
$1/\sqrt{\NN} \sim 0.001 \ll \sigma_{uc}$, and we are in the
systematics dominated regime.

For case~1 (one experiment), however, $\sigma_{uc}$ is expected to be
very small, since it only contains the time-dependent evolution of the
systematics which is different for the near and far detectors.  We
only have to worry about values larger than about $1/\sqrt{\NN}$,
which is $\simeq 0.1\%$ for a Reactor-II setup, and we neglect it for
the moment.  In the statistics dominated regime $1/\sqrt{\NN} \gtrsim
\sigma_c$, one can neglect the second term in \equ{sigmau} and one
recovers the same limiting expression as for the two-experiment
setup. For large $\NN$ (and neglecting $\sigma_{uc}$), the second term
limits the accuracy. The $1/\sqrt{\NN}$ scaling is cut off when the
condition
\begin{equation}\label{equ:scaling}
\sigma_c^2 \, \frac{(f_a - f_m)^2}{f_a^2f_m^2} \ll \frac{1}{\NN}
\end{equation}
is violated. Note that $(f_a-f_m)^2$ is a small number: Using
\equ{def} one finds $f_a-f_m \simeq \epsilon \, \langle \sin (\Delta
m^2_a L/2E)\rangle$, where $\epsilon \sim \delta_\Delta/\Delta m^2_a
\lesssim 0.1$. Furthermore, also the second factor is small, since the
fact that the experiments work at the oscillation maximum implies
$\Delta m^2_a L/2E \simeq \pi$. Using the numbers for Reactor-II, it
is straight forward to estimate that the condition in \equ{scaling} is
fulfilled, and hence, it is justified to neglect $\sigma_c$ for a
one-experiment setup. We conclude that the case~1 stays up to very
high event numbers in the statistics dominated regime, and we simulate
it in the next section by the most optimistic assumption of a
systematics-free measurement.

%%%%%%%%%%%%%%%%%%%%%%%%%%%%%%%%%%%%%%%%%%%%%%%%%%%%%%%
\section{Phenomenological results}
%%%%%%%%%%%%%%%%%%%%%%%%%%%%%%%%%%%%%%%%%%%%%%%%%%%%%%%

\begin{figure*}[t]
\begin{center}
\includegraphics[width=\textwidth]{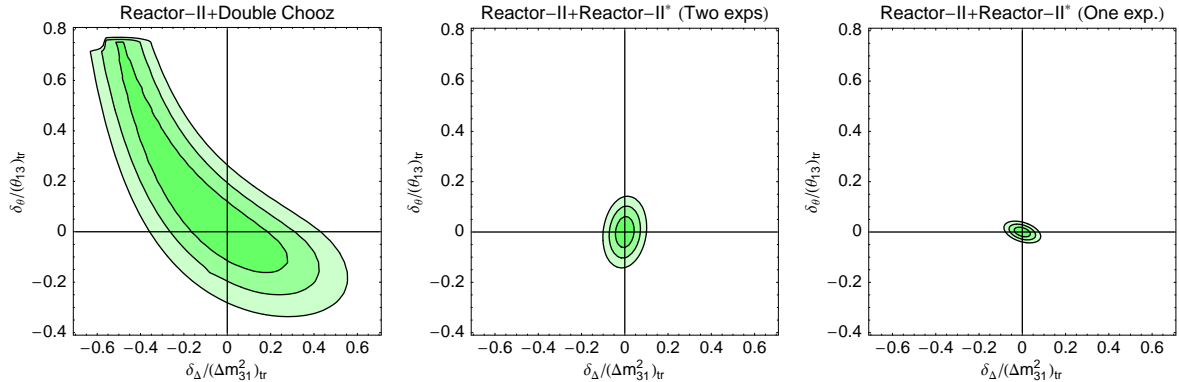}
\end{center}
\vspace*{-8ex}
\caption{\label{fig:rescomp} The combined sensitivities to
  $\delta_\Delta$ and $\delta_\theta$ (as relative changes between air
  and matter) for $\stheta_\mathrm{tr}=0.1$ at the $1\sigma$, $2
  \sigma$, and $3 \sigma$ CL (2 d.o.f) for different experiments. The
  left panel refers to Double Chooz combined with a large reactor
  experiment Reactor-II in matter, the middle panel refers to two
  different experiments Reactor-II (matter) + Reactor-II$^*$ (100~m
  matter, 1500~m air, 100~m matter), and the right panel refers to the
  optimal case with the same experiment Reactor-II (matter, phase~I) +
  Reactor-II$^*$ (phase~II), \ie, actually the material is moved. }
\end{figure*}

\begin{figure}[!tb]
\begin{center}
\includegraphics[width=0.45\textwidth]{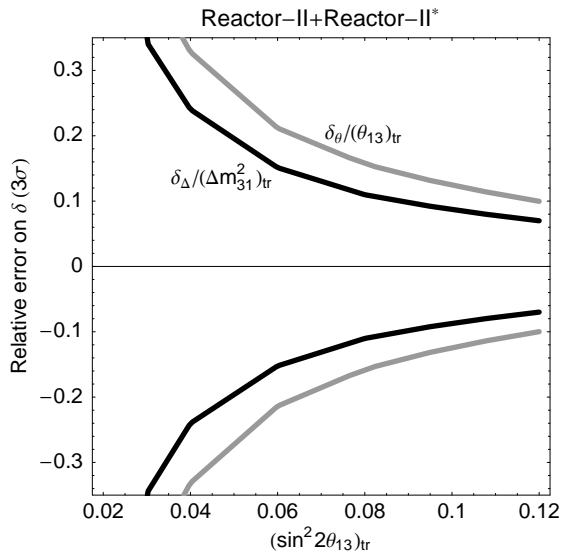}
\end{center}
\vspace*{-8ex}
\caption{\label{fig:t13dep} The sensitivities to $\delta_\Delta$ and
  $\delta_\theta$ (as relative changes between air and matter) at the
  $3 \sigma$ CL (1 d.o.f) as a function of the true $\stheta$ for two
  different experiments Reactor-II (matter) + Reactor-II$^*$ (100~m
  matter, 1500~m air, 100~m matter).}
\end{figure}

Let us first discuss the case of using experiments proposed for
measuring the standard oscillation parameters. We show in
\figu{rescomp}, left, the combined sensitivity of Double Chooz (air)
and the reactor experiment Reactor-II (matter). Obviously, there is a
strong correlation between $\delta_\Delta$ and $\delta_\theta$, which
one can understand as follows: The experiment Reactor-II measures the
matter parameters because of the much better statistics. For
$\delta_\Delta > 0$, the $\ldm$ in air can be measured by Double Chooz
by the spectral information. However, for $\delta_\Delta < 0$, the
oscillation phase becomes $\Delta_{31} \ll 1$, and the oscillation
probability in \equ{osc} can be expanded as $1-P_{\bar{e} \bar{e}}
\simeq \sin^2(2\theta_{13}) \, \Delta_{31}^2 + \hdots\,.$ This implies
that a smaller $\delta_\Delta$ (changing $\ldm$) can be directly
compensated by a larger $\delta_\theta$ (changing $\theta_{13}$),
which can be clearly seen in \figu{rescomp}, left. The final precision
for $\delta_\Delta$ and $\delta_\theta$ is only comparable to the
order of $\ldm$ and $\stheta$ themselves.

As a next step, one could think about combining two very similar
reactor experiments, one in air and one almost in matter. We show in
\figu{rescomp}, middle, the results for this setup for two experiments
with $L=1.7 \, \mathrm{km}$. Since both experiments practically have
the same spectrum, the precision on $\delta_\Delta$ and
$\delta_\theta$ is extremely improved (below 10\% at $3 \sigma$, 1
d.o.f.). The correlation between $\delta_\Delta$ and $\delta_\theta$
has almost completely disappeared, because of the long enough
baseline. We find that there is no strong baseline dependence for
$1.5~\mathrm{km} \lesssim L \lesssim 3.5~\mathrm{km}$, as can be seen
from the lower part of \Tab~\ref{tab:results}.
In \figu{t13dep} we show the dependence of the sensitivity on
$\stheta$. One observes that for $\stheta \gtrsim 0.04$, the
difference of the parameters in air and matter can be 
constrained with a precision of better than about 30\% at the
$3\sigma$ CL.\footnote{Note that due to the small values of
$\theta_{13} \sim 0.1$ a relative precision of 30\% implies the
impressive accuracy of $\sim0.03$ for $\delta_\theta$.}
Similar results can be obtained for the combination of an air Double
Chooz or Reactor-II experiment and a superbeam, such as
T2K~\cite{Itow:2001ee} or NO$\nu$A~\cite{Ayres:2004js} (\cf,
\Tab~\ref{tab:results}). In this case the sensitivity to
$\delta_\theta$ is somewhat degraded because of additional
correlations with $\deltacp$ and $\theta_{23}$ in the beam experiment.

\begin{table*}[t]
\begin{center}
\begin{tabular}{lrr}
\hline
Experiment combination &  $\delta_\Delta/(\ldm)_{\mathrm{tr}}$ & $\delta_\theta/(\theta_{13})_{\mathrm{tr}}$  \\
\hline
Double Chooz (air) + Reactor-II (matter) & 0.63 (0.29) & 0.77 (0.29) \\ 
Double Chooz (air) + NO$\nu$A (matter) &  0.75 (0.30) & $>1$ (0.34) \\ 
Reactor-II (matter) + Reactor-II$^*$   & 0.09 (0.03) & 0.12 (0.04) \\
Reactor-II (matter), transformed into Reactor-II$^*$  & {\bf 0.07 (0.02)} & {\bf 0.04 (0.01)} \\
Reactor-II (air) + NO$\nu$A (matter) & 0.08 (0.03) & 0.37 (0.23) \\
\hline
Reactor-II + Reactor-II$^*$, $L=1.0 \, \mathrm{km}$ & 0.33 (0.10) & 0.20 (0.05) \\
Reactor-II + Reactor-II$^*$, $L=2.0 \, \mathrm{km}$ & 0.08 (0.02) & 0.13 (0.04) \\
Reactor-II + Reactor-II$^*$, $L=3.0 \, \mathrm{km}$ & 0.07 (0.01) & 0.13 (0.04) \\
\hline
\end{tabular}
\end{center}
\caption{\label{tab:results}Relative sensitivities to $\delta_\Delta$
  and $\delta_\theta$ at the 3$\sigma$ (1$\sigma$) CL (1 d.o.f.) for
  different experimental setups and $\stheta_\mathrm{tr} = 0.1$. If
  the upper and lower bounds are different, the maximum value is
  given.  The definition of Reactor-II$^*$ is 100~m matter, 1500~m
  air, 100~m matter, whereas Reactor-II is in matter unless stated
  otherwise. The standard value for the baseline of Reactor-II is
  $L=1.7 \, \mathrm{km}$ unless stated otherwise.}
\end{table*}

Finally, we show in \figu{rescomp}, right, the most optimistic results
for the {\em same} experiment, \ie, using Reactor-II when 1.5~km of
the matter is removed after the phase~I run. For both phases we assume
a running time of 5 years. As discussed in the previous section, we
simulate this situation by setting all systematical errors to zero,
which leads to a further improvement of the sensitivity to
$\delta_\theta$. As highlighted in \Tab~\ref{tab:results}, such a
setup allows a test of MVNs at the level of a few per cent at
$3\sigma$~CL.

%%%%%%%%%%%%%%%%%%%%%%%%%%%%%%%%%%%%%%%%%%%%%%%%%%%%%%%%%%%%%%%%%%%%%
\section{Implications for mass-varying neutrino models}
%%%%%%%%%%%%%%%%%%%%%%%%%%%%%%%%%%%%%%%%%%%%%%%%%%%%%%%%%%%%%%%%%%%%%
\label{sec:models}

In the preceding section, we have demonstrated that reactor
experiments could provide powerful tests of any neutrino oscillation
model which is different in air and matter in the $\ldm$-$\stheta$
sector. Now we discuss some mass-varying neutrino model-dependent
aspects.

Let us first estimate the sensitivity to the elements $M_{13}$,
$M_{33}$ of the environment-dependent mass matrix in \equ{MVN}. For
simplicity we take $M_{11} = 0$. Assuming an accuracy $\epsilon \sim
\delta_\Delta/\ldm \sim \delta_\theta/\theta_{13}$, a rough estimation
gives a sensitivity of $M_{13}, M_{33} \lesssim \epsilon \sqrt{\ldm}$.
Now one can consider for example the following model for the
environment-dependent mass
matrix~\cite{Barger:2005mn,Gonzalez-Garcia:2005xu}: $M_{ij} =
\lambda^\nu_{ij} \lambda^f n_f / m_S^2$, where $f$ denotes a fermion
in the medium, $\lambda^\nu_{ij}$ ($\lambda^f$) are the Yukawa
couplings of neutrinos (the fermion $f$) to the acceleron, $n_f$ is
the number density of $f$, and $m_S$ is the acceleron mass. If we
consider only the coupling to electrons, we have $n_e \simeq 6.4\cdot
10^9$~eV$^3$ in earth matter (for $\rho=2.8 \, \mathrm{g/cm^3}$), and
we find the following order of magnitude for the sensitivity of
reactor experiments to the model parameters:
\begin{equation}
|\lambda^\nu\lambda^e|
\left(\frac{10^{-7}\:\mathrm{eV}}{m_S}\right)^2
\lesssim 2\cdot 10^{-27} \,
\left(\frac{\epsilon}{0.03}\right) \,.
\end{equation}
This number has to be compared to the value $3\cdot 10^{-28}$ obtained
in \Ref~\cite{Gonzalez-Garcia:2005xu} for the sensitivity of solar
neutrino data to MVN parameters in the 1-2 sector.  Naively one
expects that solar neutrino experiments are much more sensitive to MVN
effects than reactor experiments, since there is roughly a factor of
100 between the densities in the solar core and earth matter, and
there is another factor $\sqrt{\Delta m^2_{31}/\Delta m^2_{21}} \sim
5$, which characterizes the mass scale sensitivity of the experiments.
However, these factors are partially compensated by the high precision
of reactor experiments, $\epsilon \sim 0.03$.

In principle solar+KamLAND neutrino data have some sensitivity to
$\theta_{13}$~\cite{Fogli:2005cq,Bahcall:2004ut,Bandyopadhyay:2004da,Maltoni:2004ei}. The
above argument on the factor 100 between the densities in the sun and
the earth suggests that solar data should also be able to provide some
information on MVN effects for $\theta_{13}$. Such a constraint
requires model-dependent assumptions about the density dependence,
since one has to relate the environment-dependent values of
$\theta_{13}$ inside the sun, in earth matter, and in air.  Depending
on such model assumptions, one could have constraints on
$\delta_\theta$ from solar neutrino data, which could change the
interpretation of our results: For $\delta_\theta$ fixed to $0$,
already planned experiments in matter in combination with Double Chooz
will provide very good results for $\delta_\Delta$ (\cf, left panel of
\figu{rescomp}).

A very particular mass-varying neutrino model has been considered in
\Ref~\cite{Barger:2005mh} to explain the LSND experiment and a
potential null result in MiniBOONE by the combination of mass-varying
neutrinos and one additional sterile neutrino mixing with the active
ones in matter. Since this model predicts no oscillations 
in air for short-baseline reactor experiments,
a signal for $\stheta$ in air (for example in Double Chooz) 
would be a strong rejection of this model, \ie, $\delta_\Delta$
has to be tested only of the order of one.  
Furthermore, our setups can test any additional sterile neutrinos with
mixings different in air and matter~\cite{Barger:2005mh,Weiner:2005ac}
as follows: If the fast oscillation frequency associated with the
sterile neutrinos $\Delta M^2$ leads to oscillation effects already at
the near detector site (such as for $\Delta M^2 \gtrsim 1\,
\mathrm{eV}^2$), then both the near and far detectors will observe a
reduced overall flux. This can neither be faked by $\stheta$
nor $\ldm$ being different in matter. Thus, a sterile neutrino
contribution in matter can be measured on the level the reactor flux
is known, typically some per cent, see \eg, \Ref~\cite{Huber:2004xh}.

%%%%%%%%%%%%%%%%%%%%%%%%%%%%%%%%%%%%%%%%%%%%%%%%%%%%%%%%%%%%%%%%
\section{Summary and conclusions}
%%%%%%%%%%%%%%%%%%%%%%%%%%%%%%%%%%%%%%%%%%%%%%%%%%%%%%%%%%%%%%%%

We have considered several reactor experiment setups with baselines in
air and matter to constrain any non-standard contribution to the
Hamiltonian which is different in air and matter. In particular, we
find that new reactor experiments with near and far detectors could
provide stringent bounds for mass-varying neutrino models, which lead
to environment dependent effects in the 1-3 sector ($\ldm$ or
$\stheta$).

Non-trivial constraints can be obtained already from Double Chooz
(mainly air) combined with a matter experiment, although in this case
correlations between the parameters limit the sensitivity.  Best
results, however, are obtained if two identical setups with baselines
of at least $1.5 \, \mathrm{km}$ with substantially different material
between the near and far detectors are compared. For $\stheta \gtrsim
0.04$, deviations of oscillation parameters in matter and air can be
constrained at the level of few per cent.  Using the same experiment
and physically moving the matter between the two detectors after the
initial operation period decreases the impact of systematical errors.
In this case, the relevant systematics issue is the uncorrelated
time-dependent change between the near and far detectors, which should
be extremely small for practical purposes.
We conclude that new reactor experiments could be excellent candidates
for the test of mass-varying neutrinos provided that $\stheta$ is not
to small. 

\subsection*{Acknowledgments}

We thank P.~Huber, M.~Maltoni and R.~Zukanovich for illuminating
discussions. Furthermore, we would like to thank J.~Anjos, M.~Goodman,
and Y.~Wang for useful information.
WW would like to acknowledge support from the W.~M.~Keck Foundation
and NSF grant PHY-0503584. TS is supported by a ``Marie Curie
Intra-European Fellowship within the 6th European Community Framework
Program.''

%\bibliographystyle{apsrev}
%\bibliography{references}

\end{document}